\documentclass[a4paper,twoside]{article}
\newcommand{\affil}[1]{$^{\rm #1}$}
\usepackage[authoryear]{natbib}
\bibpunct{(}{)}{;}{a}{}{,}
\usepackage{graphicx}
\date{} 
\title{\large\bf\flushleft Co-production of light p-, s- and r-process isotopes in the high-entropy wind of 
type II supernovae}
\author{\parbox{\textwidth}{\flushleft
\vspace{-0.5cm}
{\it K. Farouqi\affil{A}, K.-L. Kratz\affil{B,C} and B. Pfeiffer\affil{B}}\\
\vspace{0.4cm}
{\small \affil{A}\,Department of Astrophysics and Astronomy, University of Chicago, Chicago, IL 60637, USA }\\
{\small \affil{B}\,Max-Planck-Institut f\"ur Chemie, Otto-Hahn-Institut, D-55128 Mainz, Germany}\\ 
{\small \affil{C}\,Email: kratz@mpch-mainz.mpg.de}}}%
\begin{document}
\twocolumn[
\begin{changemargin}{.8cm}{.5cm}
\begin{minipage}{.9\textwidth}
\vspace{-1cm}
\maketitle
%
%
\small{\bf Abstract:\\
We have performed large-scale nucleosynthesis calculations within the high-entropy-wind 
(HEW) scenario of type II supernovae. The primary aim was to constrain the conditions 
for the production of the classical "p-only" isotopes of the light trans-Fe elements. We find, 
however, that for electron fractions in the range 0.458 $\le$ Y$_e$ $\le$ 0.478, sizeable 
abundances of p-, s- and r-process nuclei between $^{64}$Zn and $^{98}$Ru are coproduced 
in the HEW at low entropies (S $\le$ 100) by a primary charged-particle process after an 
$\alpha$-rich freezeout. With the above Y$_e$ -- S correlation, most of the predicted isotopic 
abundance ratios within a given element (e.g. $^{64}$Zn(p)/$^{70}$Zn(r) or $^{92}$Mo(p)/$^{94}$Mo(p)), 
as well as of neighboring elements (e.g. $^{70}$Ge(s+p)/$^{74}$Se(p) or $^{74}$Se(p)/$^{78}$Kr(p)) 
agree with the observed Solar-System ratios. Taking the Mo isotopic chain as a particularly 
challenging example, we show that our HEW model can account for the production of all 7 
stable isotopes, from "p-only" $^{92}$Mo, via "s-only" $^{96}$Mo up to "r-only" $^{100}$Mo. 
Furthermore, our model is able to reproduce the isotopic composition of Mo in presolar SiC 
X-grains.} 
\medskip
\medskip
\end{minipage}
\end{changemargin}
]
\small
\section{Introduction}
The origin of the stable isotopes of the light trans-Fe elements in the Solar System (SS) has been a fascinating area for nuclear astrophysicists over more than 50 years. It is commonly believed that these elements (between Zn (Z=30) and about Cd (Z=48)) are produced by varying contributions from three historical nucleosynthesis processes, (i) the "p-process" (see, {\it e.g.} \citet{b2fh,Arnould76,WoosleyHoward}), (ii) the "weak s-process" (see, {\it e.g.} 
\citet{Clayton,Kappeler82,Kappeler89}), and (iii) the "weak r-process" (see, {\it e.g.} \citet{Seeger,Hillebrandt,Cowan91,Kratz93}). \\
Apart from the SS {\bf isotopic} abundances \citep{Lodders}, astronomical observations in recent years of {\bf elemental} abundances in ultra-metal-poor (UMP) halo stars (see, {\it e.g.} \citet{Barklem,Francois,Mashonkina}) revived and intensified interest in the nucleosynthesis of these elements, and have motivated various theoretical studies with increasing realism (see, {\it e.g.} \citet{Hoffman96,Rauscher02,Travaglio,Frohlich,Far08a,Far08b,Kratz08,Pignatari,Wanajo,Far09}). In addition, measurements of the {\bf isotopic} compositions of trans-Fe elements in presolar SiC grains of type X (see, {\it e.g.} \citet{Pellin2000,Pellin2006,Marhas2007}), motivated a suggestion for a fourth process contributing to these isotopes, i.e. a neutron burst in the shocked He shell of a supernova \citep{Meyer}. However, even those recent models have major shortcomings the one or other way. In particular, as in the older models, still none of the presently favored astrophysical scenarios produces sufficiently high abundances of all p-nuclei of Zn (Z=30) to Ru (Z=44), and all models 
seem to be unable to reproduce the SS abundance ratio of the two highly abundant p-isotopes $^{92}$Mo and $^{94}$Mo \citep{Lodders}. \\ 
The high-entropy wind (HEW) of core-collapse type II supernovae (SN II) may offer a solution to the above problems by producing the light trans-Fe elements by a primary charged-particle ($\alpha$-) process. This nucleosynthesis process seems to be largely uncorrelated (Zn -- Nb) or weakly correlated (Mo -- Cd) with the "main" r-process 
at and beyond the A$\simeq$130 abundance peak
\citep{Wooshoff,Qian07,Hoffman08b,Far08a,Far08b,Kratz08,Far09}, as is indicated by 
recent astronomical observations. In the present paper, we describe under which conditions of electron abundance (Y$_e$=Z/A), entropy (S$\sim$T$\rho$ [k$_b$/baryon]) and a selected expansion speed of the ejecta (V$_{exp}$=7500 [km/s]), the HEW scenario can co-produce p-, s- and r-process isotopes of the trans-Fe elements between Zn (Z=30) and Ru(Z=44). We present absolute yields in units of M$_{\odot}$ for a pure charged-particle ($\alpha$-) process for the choice of three typical electron abundances of Y$_e$=0.450, 0.470 and 0.490. \\
Furthermore, we show for a number of selected cases that the predicted isotopic 
abundance ratios within a given element (e.g. $^{64}$Zn(p)/$^{70}$Zn(r)), as well as of 
neighboring elements (e.g. $^{70}$Ge(s+p)/$^{74}$Se(p)) agree well with the observed 
SS abundance ratios. However, as in other models, we cannot completely avoid certain 
isotopic over-abundances (e.g. for $^{88}$Sr and $^{90}$Zr) or under-abundances (e.g. 
for $^{96,98}$Ru).  \\    
Taking the Mo isotopic chain as a particularly challenging example, we show that the 
$\alpha$-component of our HEW model can account for the production of all seven stable 
isotopes, from "p-only" $^{92}$Mo, via "s-only" $^{96}$Mo up to "r-only" $^{100}$Mo. 
Finally, we indicate that our model is also able to reproduce the isotopic composition of Mo 
in presolar SiC X-grains, recently measured by \citep{Pellin2000,Pellin2006}.  \\   
Unlike the "neutron-burst" model of \citep{Meyer} and the "$\gamma$-process" in 
the pre-SN and SN models of \citep{Rauscher02} (which both start from an initial SS "seed" 
composition); however similar to the "neutrino-wind" model of \citep{Hoffman96} and 
the recent "electron- capture SN" model of \citep{Wanajo}, the $\alpha$-component 
of our HEW scenario is a {\bf primary} process. This means that the low-S production of 
all light trans-Fe isotopes (classical p-, s- and r) does not require any assumptions about the 
initial composition of the SN progenitor star.    

\section{Calculations and Results}
The concept of a high-entropy wind (HEW) arises from considerations of the newly born proto-neutron star in core-collapse supernovae. In this scenario, the late neutrinos interact with matter of the outermost neutron-star layers, leading to moderately neutron-rich ejecta with high entropies (see, {\it e.g.}
\citet{Woosley94,Hoffman96,Freiburg}; and for recent publications, see, {\it e.g.} 
\citet{Heger,Hoffman08}, and references therein). As in \citep{Far08a,Far08b,Far09},  in the calculations presented here we follow the description of adiabatically expanding mass zones as previously utilized in \citep{Freiburg}. The nucleosynthesis calculations up to the charged-particle freezeout were performed with the latest Basel code, but without taking into account neutrino-nucleon / nucleus interactions. Neutrino-accelerated nucleosynthesis, the so-called ($\nu$p)-process (see, {\it e.g.} \citet{Frohlich}) produces proton-rich matter and drives the nuclear flow into the "light" trans-Fe region, contributing considerably to the production of elements up to the Zn (Z=30) -- Ge (Z=32) region, but then presumably fading out quickly in the Se (Z=34) -- Rb (Z=37) region. \citep{Frohlich} predict that the $\nu$p-process would also efficiently synthesize p-nuclei between Sr (Z=38) and Pd (Z=46). However, we believe that with more realistic neutrino fluxes, the $\nu$p-process will not contribute significantly to the "heavier" trans-Fe elements. Therefore, we assume that these elements are primarily produced by the charged-particle ($\alpha$-) component of the HEW. \\ 
Reaction rates for the HEW model were taken from the Hauser-Feshbach model NON-SMOKER 
\citep{Rauscher00,Rauscher07}. The subsequent parameterized "r-process" network calculations use updated experimental and theoretical nuclear-physics input on masses and $\beta$-decay properties, as outlined in \citep{Kratz07a} and used in our earlier "waiting-point" calculations (see, {\it e.g.} \citet{Kratz93,Kratz07b}). \\ 
After charged-particle ($\alpha$-) freezeout, the expanding and eventually ejected mass zones have different initial entropies (S), so that the overall explosion represents a superposition of entropies correlated with different electron abundances (Y$_e$), different ratios of free neutrons to "seed" nuclei (Y$_n$/Y$_{seed}$), and eventually different expansion velocities (V$_{exp}$) as well \citep{Far08a,Far08b,Kratz08}.
If one assumes that equal amounts of ejected material per S-interval are contributing, the sum of these abundance fractions is weighted according to the resulting Y$_{seed}$ as a function of S (see, {\it e.g.} Fig. 1 in \citet{Far08a} and / or \citet{Far09}). From this parameter study, we have found that the HEW predicts -- at least -- two clearly different nucleosynthesis modes. For low entropies (e.g. 5$\le$S$\le$110 at Y$_e$=0.450), the concentration of free neutrons is negligible. Hence, the nucleosynthesis in this S-range is definitely not a neutron-capture process but rather a charged-particle ($\alpha$-) process. For higher entropies, the Y$_n$/Y$_{seed}$ ratios are increasing smoothly, resulting in a neutron-capture component resembling a classical "weak" r-process followed by the classical "main" r-process, which produces the heavy nuclei up to the Th, U actinide region. \\

In our previous papers \citep{Far08a,Far08b,Kratz08,Far09}, we have compared our HEW model results to the classical SS {\bf isotopic} 
"r-residuals" (N$_{r,\odot}$= N$_\odot$--N$_s$; (see, e.g. \citet{Kappeler89,Arlandini}) and to recent {\bf elemental} abundances in 
UMP halo stars (see, e.g. \citet{Barklem,Cowan06,Mashonkina}). In these papers we have demonstrated that a 
superposition of entropies in the full range of 5$\le$S$\le$300 for a single electron 
fraction of Y$_e$=0.450 was able to very well reproduce the SS "main" r-process 
distribution in the mass range 120$\le$A$\le$209 (see, e.g. Fig. 2 in \citet{Kratz08} 
and / or \citet{Far09}). However, it also became obvious that this 
way of weighting the S-components for a single Y$_e$ value did not fit the classical 
"r-residuals" in the region between the Fe-group and the rising wing of the A$\simeq$130 
peak, neither for the SS isotopic abundances, nor for the element abundances in the majority of the 
UMP stars. In particular for the light trans-Fe elements of Zn (Z=30) to Rb (Z=37) 
in these stars, it was evident that the HEW model predicts much too low Z-abundances 
(see, e.g. Fig. 3 in \citet{Kratz08} and / or \citet{Far09}).  \\    
 There have been several suggestions to explain the abundances in this mass region 
with a multiplicity of nucleosynthesis processes. The first authors who recognized this 
possibility were \citep{Hoffman96} with their neutrino-driven wind model (see 
also \citet{Hoffman97}). In this parameter study, with the restriction 
to a single low entropy of S$\simeq$50 and a variety of electron fractions in the range 
0.46$\le$Y$_e$$\le$ 0.50, they noted that {\it "the r-process and some light p-process 
nuclei may be coproduced"} in a primary charged-particle process. Later, after the first 
measurements of some elements in the trans-Fe region of UMP halo stars had become 
available, a light element primary process "LEPP" was invoked by \citep{Travaglio}, 
qualitatively related to s-like neutron captures. A recent more quantitative alternative to 
such a neutron-capture scenario could be a strong secondary s-process with a primary 
($\alpha$,n) neutron source in massive stars at low metallicities, as suggested by 
\citep{Pignatari}. On the other hand, in the latest revised version of the phenomenological 
"LEGO" model of \citep{Qian07,Wasserburg}, following the basic arguments of \citep{Hoffman96},
consider the trans-Fe elements to be dominantly produced by charged-particle 
reactions. \\  
 As already discussed in previous papers (see, e.g.\citet{Far08a,Far08b,Kratz08,Far09})
our HEW approach with the above parameter choices of 
individual Y$_e$ and V$_{exp}$ and superpositions of S did neither fully support the above 
"LEPP" nor the "LEGO" idea. We have shown that the low-S region (S$\le$100 for a 
Y$_e$=0.45), is indeed a pure charged-particle process, producing the lighter trans-Fe 
elements up to about Nb (Z=41). This is in agreement with the initial ideas of \citep{Hoffman96}
and the later "LEGO" approach, but in disagreement with the "LEPP" idea. From Mo 
(Z=42) on upwards, however, our HEW model predicts smoothly increasing fractions of 
neutron-capture material, now in qualitative agreement with the "LEPP" approach, but in 
disagreement with the "LEGO" picture.  \\
After having focussed on {\it elemental} abundances in the past, in this paper 
we want to discuss first HEW results on {\it isotopic} abundances of the 
trans-Fe 
elements between Zn (Z=30) and Ru (Z=44), in particular their decomposition 
into the respective fractions of the historical p-, s- and r-process nuclei. We 
start by presenting in Table~\ref{table1} the isotopic abundances between $^{64}$Zn and 
$^{98}$Ru in units of solar masses M$_{\odot}$ for three typical neutrino-wind 
conditions. For a selected expansion velocity of V$_{exp}$=7500 [km/s], we 
consider for the other two correlated parameters (i) a "neutron-rich" component 
with Y$_e$=0.450 and an entropy superposition of 5$\le$ S $\le$100 (S$\le$100), 
(ii) a "proton-rich" component with Y$_e$=0.490 and a superposition of S$\le$150, 
and (iii) a "medium" component with Y$_e$=0.470 and a superposition of S$\le$120, 
where the maximum entropy for each value of Y$_e$ is defined by a neutron-to-seed 
ratio Y$_n$/Y$_{seed}$=1.0.
With these parameter choices only the charged-particle ($\alpha$-) component of the total HEW abundances is considered. \\
\begin{table*}[t]
\begin{center}
\caption{Yields of stable isotopes (in units of M$_\odot$) for the $\alpha$-component of the HEW}
\label{table1}
\begin{tabular}{rrrrrr}
\hline 
\multicolumn{2}{c}{Y$_e$=0.450} &\multicolumn{2}{c}{Y$_e$=0.470}& \multicolumn{2}{c}{Y$_e$=0.490} \\
Isotope & Yield [$M_{\odot}$] & Isotope & Yield [$M_{\odot}$] & Isotope & Yield [$M_{\odot}$] \\
\hline
$^{ 64}$Zn&.34E-06&$^{ 64}$Zn&.56E-04&$^{ 64}$Zn&.96E-05\\
$^{ 66}$Zn&.35E-04&$^{ 66}$Zn&.55E-04&$^{ 66}$Zn&.61E-05\\
$^{ 67}$Zn&.51E-06&$^{ 67}$Zn&.71E-06&$^{ 67}$Zn&.82E-07\\
$^{ 68}$Zn&.92E-05&$^{ 68}$Zn&.20E-05&$^{ 68}$Zn&.41E-06\\
$^{ 70}$Zn&.19E-07&$^{ 70}$Zn&.10E-07&$^{ 70}$Zn&.71E-08\\
$^{ 69}$Ga&.49E-06&$^{ 69}$Ga&.38E-06&$^{ 69}$Ga&.51E-07\\
$^{ 71}$Ga&.11E-06&$^{ 71}$Ga&.11E-06&$^{ 71}$Ga&.16E-07\\
$^{ 70}$Ge&.10E-05&$^{ 70}$Ge&.89E-05&$^{ 70}$Ge&.97E-06\\
$^{ 72}$Ge&.56E-05&$^{ 72}$Ge&.23E-05&$^{ 72}$Ge&.30E-06\\
$^{ 73}$Ge&.81E-07&$^{ 73}$Ge&.89E-07&$^{ 73}$Ge&.16E-07\\
$^{ 74}$Ge&.71E-06&$^{ 74}$Ge&.54E-07&$^{ 74}$Ge&.13E-07\\
$^{ 76}$Ge&.23E-07&$^{ 76}$Ge&.20E-07&$^{ 76}$Ge&.14E-07\\
$^{ 75}$As&.17E-06&$^{ 75}$As&.77E-07&$^{ 75}$As&.16E-07\\
$^{ 74}$Se&.61E-08&$^{ 74}$Se&.53E-06&$^{ 74}$Se&.53E-07\\
$^{ 76}$Se&.64E-06&$^{ 76}$Se&.17E-05&$^{ 76}$Se&.21E-06\\
$^{ 77}$Se&.63E-07&$^{ 77}$Se&.59E-07&$^{ 77}$Se&.11E-07\\
$^{ 78}$Se&.15E-05&$^{ 78}$Se&.34E-06&$^{ 78}$Se&.51E-07\\
$^{ 80}$Se&.15E-06&$^{ 80}$Se&.37E-07&$^{ 80}$Se&.17E-07\\
$^{ 82}$Se&.13E-06&$^{ 82}$Se&.84E-07&$^{ 82}$Se&.28E-07\\
$^{ 79}$Br&.10E-06&$^{ 79}$Br&.85E-07&$^{ 79}$Br&.18E-07\\
$^{ 81}$Br&.17E-06&$^{ 81}$Br&.48E-07&$^{ 81}$Br&.16E-07\\
$^{ 78}$Kr&.45E-10&$^{ 78}$Kr&.40E-07&$^{ 78}$Kr&.39E-08\\
$^{ 80}$Kr&.95E-08&$^{ 80}$Kr&.23E-06&$^{ 80}$Kr&.25E-07\\
$^{ 82}$Kr&.14E-06&$^{ 82}$Kr&.28E-06&$^{ 82}$Kr&.37E-07\\
$^{ 83}$Kr&.86E-07&$^{ 83}$Kr&.59E-07&$^{ 83}$Kr&.24E-07\\
$^{ 84}$Kr&.17E-05&$^{ 84}$Kr&.60E-06&$^{ 84}$Kr&.12E-06\\
$^{ 86}$Kr&.18E-04&$^{ 86}$Kr&.82E-05&$^{ 86}$Kr&.17E-05\\
$^{ 85}$Rb&.14E-05&$^{ 85}$Rb&.64E-06&$^{ 85}$Rb&.14E-06\\
$^{ 87}$Rb&.45E-05&$^{ 87}$Rb&.22E-05&$^{ 87}$Rb&.56E-06\\
$^{ 84}$Sr&.20E-10&$^{ 84}$Sr&.12E-07&$^{ 84}$Sr&.12E-08\\
$^{ 86}$Sr&.99E-07&$^{ 86}$Sr&.14E-06&$^{ 86}$Sr&.21E-07\\
$^{ 87}$Sr&.16E-06&$^{ 87}$Sr&.85E-07&$^{ 87}$Sr&.15E-07\\
$^{ 88}$Sr&.12E-03&$^{ 88}$Sr&.31E-04&$^{ 88}$Sr&.65E-05\\
$^{ 89}$Y &.18E-04&$^{ 89}$Y &.92E-05&$^{ 89}$Y &.20E-05\\
$^{ 90}$Zr&.23E-04&$^{ 90}$Zr&.28E-04&$^{ 90}$Zr&.56E-05\\
$^{ 91}$Zr&.46E-05&$^{ 91}$Zr&.22E-05&$^{ 91}$Zr&.51E-06\\
$^{ 92}$Zr&.71E-05&$^{ 92}$Zr&.31E-05&$^{ 92}$Zr&.58E-06\\
$^{ 94}$Zr&.98E-05&$^{ 94}$Zr&.39E-05&$^{ 94}$Zr&.56E-06\\
$^{ 96}$Zr&.26E-05&$^{ 96}$Zr&.86E-06&$^{ 96}$Zr&.97E-07\\
$^{ 93}$Nb&.38E-05&$^{ 93}$Nb&.15E-05&$^{ 93}$Nb&.23E-06\\
$^{ 92}$Mo&.34E-09&$^{ 92}$Mo&.26E-07&$^{ 92}$Mo&.44E-08\\
$^{ 94}$Mo&.55E-09&$^{ 94}$Mo&.20E-08&$^{ 94}$Mo&.46E-09\\
$^{ 95}$Mo&.11E-05&$^{ 95}$Mo&.43E-06&$^{ 95}$Mo&.76E-07\\
$^{ 96}$Mo&.54E-10&$^{ 96}$Mo&.61E-10&$^{ 96}$Mo&.33E-10\\
$^{ 97}$Mo&.32E-06&$^{ 97}$Mo&.13E-06&$^{ 97}$Mo&.16E-07\\
$^{ 98}$Mo&.78E-06&$^{ 98}$Mo&.13E-06&$^{ 98}$Mo&.97E-08\\
$^{100}$Mo&.35E-06&$^{100}$Mo&.11E-06&$^{100}$Mo&.15E-07\\
$^{ 96}$Ru&$<$E-15&$^{ 96}$Ru&.63E-12&$^{ 96}$Ru&.98E-13\\
$^{ 98}$Ru&$<$E-15&$^{ 98}$Ru&.25E-12&$^{ 98}$Ru&.45E-13\\
$^{ 99}$Ru&.86E-07&$^{ 99}$Ru&.36E-07&$^{ 99}$Ru&.17E-08\\
$^{100}$Ru&.54E-14&$^{100}$Ru&.91E-14&$^{100}$Ru&.54E-14\\
$^{101}$Ru&.21E-06&$^{101}$Ru&.70E-07&$^{101}$Ru&.38E-08\\
$^{102}$Ru&.59E-06&$^{102}$Ru&.18E-06&$^{102}$Ru&.12E-07\\
$^{104}$Ru&.65E-06&$^{104}$Ru&.13E-06&$^{104}$Ru&.26E-08\\
\hline
\end{tabular}
\medskip\\
\end{center}
\end{table*} 
Several conclusions can be drawn from our detailed HEW model calculations in the 
total Y$_e$ -- S -- V$_{exp}$ parameter range.
The first one is, that the overall yields of the 
light trans-Fe elements decrease with increasing Y$_e$. When considering the total Z-region 
between Fe and Cd, the produced $\alpha$-yields are about 4.0x10$^{-3}$ M$_{\odot}$ for 
Y$_e$=0.450 and 2.7x10$^{-3}$ M$_{\odot}$ for Y$_e$=0.490, respectively. 
These abundances can be compared with the corresponding neutron-capture r-process yields 
(for higher entropies with corresponding Y$_n$/Y$_{seed}>$1.0) of 3.4x10$^{-4}$ M$_{\odot}$ for 
Y$_e$=0.450 and 4.3x10$^{-5}$ M$_{\odot}$ for Y$_e$=0.490, respectively. The second observation is, 
that in the range 0.450$\le$Y$_e$$\le$0.480 the relative isotopic abundances of the trans-Fe elements 
are shifted towards the lighter stable nuclides favoring s-iotopes, or even to the proton-rich 
side then favoring p-isotopes. For higher electron abundances up to Y$_e$=0.498, the trend 
becomes slightly reverse.\\
Furthermore, for the range 0.460$\le$Y$_e$$\le$0.490 the HEW low-entropy charged- 
particle ($\alpha$-) process produces the lightest isotopes of all even-Z isotopes between 
Fe (Z=26) and Ru (Z=44), where all p-nuclei are involved. Above Ru (Z=44), the 
abundance fractions of the HEW $\alpha$-component become negligible compared to 
the now dominating neutron-capture "weak" r-process. Hence, sizeable isotopic yields 
for Pd (Z=46) and Cd (Z=48) are only produced for the heavier isotopes $^{105}$Pd 
and $^{111}$Cd and beyond, respectively.  \\    
For a more quantitative consideration, let us choose as typical examples the  
HEW $\alpha$-process yield ratios of the two lightest isotopes of Zn, (i.e. 
$^{64}$Zn/$^{66}$Zn), Sr, (i.e. $^{84}$Sr/$^{86}$Sr), and Mo, ( i.e. 
$^{92}$Mo/$^{94}$Mo) (see Table~\ref{table1}). 
As can be deduced from Table~\ref{table1}, the predicted yield ratio of 
$^{64}$Zn/$^{66}$Zn varies between 9.5x10$^{-3}$ for Y$_e$=0.450, 1.02 for Y$_e$=0.470 
and 1.57 for Y$_e$=0.490, where the isotopic ratio for "moderate" Y$_e$ values slightly above
0.47 seem to agree best with the measured SS yield ratio of 1.74 of \citep{Lodders}. The dominant effect in the 
HEW ratios comes from the strong increase of the $^{64}$Zn yield by roughly two 
orders of magnitude in the 0.45$\le$Y$_e$$\le$0.47 range, whereas the corresponding 
change of the $^{66}$Zn abundance is only a factor 2. A similar picture is obtained 
for the abundance ratio of $^{84}$Sr/$^{86}$Sr, resulting in 2.0x10$^{-4}$ for Y$_e$=0.450, 
8.9x10$^{-2}$ for Y$_e$=0.470 and 5.7x10$^{-2}$ for Y$_e$=0.490. Here, the best agreement with 
the SS abundance ratio of 5.66x10$^{-2}$ seems to be reached for a rather proton-rich 
Y$_e$ scenario. Finally, the predicted abundance ratio of the two "p-only" isotopes 
$^{92}$Mo and $^{94}$Mo vary between 0.62 for Y$_e$=0.450, 13 for Y$_e$=0.470 
and 9.6 for Y$_e$=0.490. In this case, none of the above HEW ratios agrees with the 
SS abundance ratio of 1.60.  \\ 
 
From the above results deduced from Table~\ref{table1}, we see that with this rather coarse HEW 
parameterization we do not obtain a consistent picture for the whole trans-Fe region. 
This seems to confirm all earlier and recent attempts which also were not able to 
obtain a satisfactory overall reproduction of the SS abundances in the region of the 
light trans-Fe p-nuclei the one or other way (see, e.g. 
\citet{Hoffman96,Hoffman97,Meyer,Rauscher02,Hoffman08,Bazin,Pignatari,Fisker,Wanajo}). 
Therefore, 
two further refinements of our model may be necessary to improve the situation obtained 
so far: (i) a finer grid of {\it individual} values of Y$_e$ in the range 0.450$\le$Y$_e$$\le$0.490, 
and / or (ii) a {\it superposition} of Y$_e$ values with model-inherent weighting of the 
respective HEW yields.  \\  
 We have followed both approaches successively. And, already with a finer grid of single 
Y$_e$ trajectories and the previously applied S-superposition up to S$_{max}$ where 
Y$_n$/Y$_{seed}$=1.0, in a large number of cases we obtain isotopic abundance ratios 
for a specific element and for nuclides of neighboring elements which are in better agreement 
with the SS values. However, as shown for example by \citep{Burrows}, even more 
realistic should be an additional weighted superposition of Y$_e$ trajectories over certain ranges. In the following, in 
Table~\ref{table2} we present our first HEW results from the second approach with correlated superpositions 
of S and Y$_e$ on selected  isotopic abundance ratios. In our attempt to obtain a consistent  
overall reproduction of the SS values, we have determinded the optimum ranges 
of the two astrophysical parameters for the later model-inherently weighted superposition. As 
already mentioned before, we have restricted our study to the full entropy ranges (S$\le$100 for 
Y$_e$=0.45, S$\le$120 for Y$_e$=0.47 and S$\le$150 for Y$_e$=0.49) responsible for a pure 
charged-particle ($\alpha$-) process. Within these S-ranges, a rather constant range of 
0.458$\le$Y$_e$$\le$0.474 has been obtained for the whole mass region from Zn up to Ru. With 
this optimized, correlated S -- Y$_e$ superposition, we have compared our predictions with the 
three selected nucleosynthesis models of \citep{Hoffman96}, \citep{Rauscher02} and 
\citep{Wanajo}. \\ 
\begin{table*}[t]
\begin{center}
\caption{Selected isotopic abundance ratios of light trans-Fe elements between Zn (Z=30) and 
Ru (Z=44). Comparison of Solar-System values \citep{Lodders} with the predictions of our HEW 
charged-particle ($\alpha$-) component (see, e.g. \citet{Kratz08,Far09} and 
previous results from three different nucleosynthesis scenarios, (i) the early SN neutrino-wind 
ejecta (Y$_e$=0.465) by  \citep{Hoffman96}, (ii) the $\gamma$-process in s-processed massive 
stars (15 M$_{\odot}$ model) by \citep{Rauscher02}, and (iii) the electron-capture (EC) SNe in 
asymptotic giant branch stars with an O-Ne-Mg core (ST model) by \citep{Wanajo}.
}
\label{table2}
\begin{tabular}{rrrrrr}
\hline
&&\multicolumn{4}{c}{Isotopic abundance ratios} \\
Isotope pairs & Solar System & This work &  Neutrino-wind & $\gamma$-process & EC SN \\
(nucleosynth. origin) &&&(1996)&(2002)&(2009) \\
\hline
$^{64}$Zn(p)/$^{70}$Zn(r)   &   78.4     &       79.4      &            /           &        10.5          &             6.6 E+7  \\ 
$^{64}$Zn(p)/$^{70}$Ge(s,p) &  23.3      &      13.6       &         0.39           &              8.63         &               7.7 \\ 
$^{70}$Ge(s,p)/$^{76}$Ge(r) &    2.84    &        4.61     &          /             &                2.53       &               2.8 E+9 \\ 
$^{70}$Ge(s,p)/$^{74}$Se(p) &  40.1      &      41.1       &    55.2              &             30.9          &              16.2 \\ 
$^{74}$Se(p)/$^{76}$Se(s)   & 9.42 E-2   &  9.09 E-2       &      /             &                 0.128     &                 0.567 \\
$^{74}$Se(p)/$^{82}$Se(r)   &    0.101   &       0.113     &       /            &                  0.120    &                6.1 E+9  \\
$^{74}$Se(p)/$^{78}$Kr(p)   &    2.90    &       11.0      &      41.8          &                 10.9      &                    7.27 \\ 
$^{78}$Kr(p)/$^{80}$Kr(p,s) &    0.156   &         0.156   &         /          &                 7.40 E-2  &                   0.245 \\
$^{78}$Kr(p)/$^{82}$Kr(s)    &  3.11 E-2 &    2.92 E-2   &         /          &                  1.97 E-2  &                   0.654 \\
$^{78}$Kr(p)/$^{86}$Kr(r,s)  &  2.11 E-2 &    7.9 E-4     &        /           &                  5.8 E-3   &                 5.7 E+4 \\
$^{78}$Kr(p)/$^{84}$Sr(p)    &   1.52     &         2.77    &       6.94       &                  1.83        &                  2.28  \\
$^{84}$Sr(p)/$^{86}$Sr(s)    &  5.66 E-2 &    4.00 E-2   &        /           &                 4.05 E-2   &                  0.240  \\
$^{84}$Sr(p)/$^{90}$Zr(s,r)   & 2.25 E-2 &    1.3 E-4     &        /           &                 2.13 E-2   &                2.6 E-3  \\
$^{84}$Sr(p)/$^{92}$Mo(p)   &   0.340   &        0.344   &      3.1 E-2    &                  0.467      &                   0.442 \\
$^{90}$Zr(s,r)/$^{96}$Zr(r,s) &  18.4      &       5.56       &      /             &              10.4            &              $>$ E+20 \\
$^{90}$Zr(s,r)/$^{92}$Mo(p)  &  15.1      &    2.2 E+3    &     2.6 E+3   &                22.0          &                172  \\
$^{92}$Mo(p)/$^{94}$Mo(p)  &    1.60    &       1.86      &       3.31       &                 1.55         &                 49.4  \\    
$^{92}$Mo(p)/$^{96}$Ru(p)  &      3.67  &       3.0 E+4 &        2.6 E+4 *)  &               3.48       &                2.7 E+4 \\
$^{96}$Ru(p)/$^{98}$Ru(p)   &    2.97    &       2.57      &    27.0 *)         &              2.54          &                 9.06 \\
\hline
\end{tabular}
\end{center}
*) Average ratio deduced from the abundances given in Table 4 of \cite{Hoffman96}, where a 
quasi-equilibrium of $^{92,94}$Mo and $^{96,98}$Ru with $^{90}$Zr was assumed.     \\
\end{table*}
It is evident from our Table~\ref{table2}, that -- as attempted by our above parameter "finetuning" -- in most 
of the cases, the isotopic abundance ratios of our HEW $\alpha$-component agree quite well with 
the SS values. Exceptions are indicated by (i) the "step" of about a factor 4 for the two p-nuclei 
$^{74}$Se/$^{78}$Kr, (ii) the well known local over-abundances in the N=50 region for Sr and Zr 
isotopes, and (iii) the low abundances of the two p-isotopes of Ru, their abundance ratio, however, 
again agreing with the SS ratio. Nevertheless, compared to the other three models, our low-S HEW 
$\alpha$-process gives the best overall agreement with the SS isotopic abundance ratios. \\ 
In the early \citep{Hoffman96} parameter study on the {\it "production of the light p-nuclei in 
neutrino-driven winds"}, who use a single entropy of S$\simeq$50 and individual electron 
fractions in the range 0.46$\le$Y$_e$$\le$0.50, the authors have noted for the first time that their 
{\it "new kind of p-process"} is primary and that {\it "the r-process and some light p-process nuclei 
may be coproduced"}. With respect to Mo and Ru, \citep{Hoffman96} conclude that $^{92}$Mo is made 
in quasi-equilibrium with $^{90}$Zr, whereas {\it "the origins of $^{94}$Mo and $^{96,98}$Ru 
remain a mystery"}. For our comparison, we have deduced the available abundance ratios from 
their Fig. 3 (for Y$_e$=0.460) and their Table 4. It is evident from our Table~\ref{table2} that, maybe except 
$^{70}$Ge/$^{74}$Se and $^{92}$Mo/$^{94}$Mo, there is no agreement between the $\nu$-wind model and  the SS values for 
all other abundance ratios of the light p-isotopes between Zn and Ru.  \\ 
\citep{Rauscher02} have presented  detailed nucleosynthesis calculations in massive stars from the 
onset of central H-burning through explosion of SN-II for Population I stars of 15$\le$M$_\odot$$\le$25. 
They find that in some stars, most of the p-nuclei can be produced in the convective O-burning shell 
prior to collapse, whereas others are made only in the explosion. Again, with respect to Mo and Ru, 
the authors point out that {\it "serious deficiencies still exist in all cases for the p-isotopes of Mo and 
Ru"}. For our comparison, we have chosen their 15 M$_\odot$ model S15 which starts with an initial 
SS "seed" composition, which is further modulated by an s-process and finally by a 
$\gamma$-process. As can be seen from our Table~\ref{table2}, for the majority of the selected isotopic 
abundances, the above model gives quite good agreement with the SS values. Similar to our HEW 
model, a large abundance "step" is observed between Se and Kr, as well as the strong over-abundances 
in the N=50 Sr -- Zr region. In addition, and in contrast to our HEW approach, the SS isotopic ratios in 
the Zn -- Ge region cannot be reproduced.  \\   
 A very recent state-of-the-art hydrodynamical simulation on the {\it "nucleosynthesis in electron-capture (EC)
supernovae of asymptotic giant branch stars with an O--Ne--Mg core"} has been performed by \citep{Wanajo}. 
For electron fractions in the range 0.464$\le$Y$_e$$\le$0.470 the authors obtain large 
productions of light p-nuclei between $^{64}$Zn and $^{92}$Mo. The correlated significant overproduction 
of $^{90}$Zr can be avoided in their model by {\it "boosting Y$_e$ to 0.480"}. Another interesting result is 
that in their EC SN scenario obviously {\it "the $\nu$p-process does not play any role in producing the 
p-nuclei"}. For comparison with our results and the SS values, we have chosen the abundances from
 their unmodified model ST. As can be seen from our Table~\ref{table2}, agreement with the SS isotopic abundances is 
only obtained in a few cases. The most evident discrepancies obviously occur for the all neutron-rich 
(r-process) nuclides of Zn up to Zr, which are orders of magnitude under-produced relative to their 
p-isotopes. Also the SS value of $^{92}$Mo/$^{94}$Mo cannot be reproduced.  \\  
 
Finally, we want to discuss explicitly the abundances of the Z=42 Mo isotopes predicted by the 
$\alpha$-component of our HEW model. There are several reasons for chosing this isotopic chain. In the light trans-Fe region, besides Ru (Z=44), Mo has with 7 nuclides the longest sequence of stable isotopes, from the two light 
"p-only" nuclei $^{92}$Mo and $^{94}$Mo (with their unusually high SS fractions of 14.84$\%$ and 9.25$\%$, respectively), via the intermediate-mass "s-only" isotope $^{96}$Mo (16.68$\%$), up to the 
"r-only" nuclide 
$^{100}$Mo (9.63$\%$) \citep{Lodders}; the remaining isotopes $^{95,97,98}$Mo have mixed s + r origin. Therefore, it is of special interest to check whether our HEW 
model can in principle account for the coproduction of all 7 stable Mo isotopes, 
and which abundance fractions relative to the SS values can be formed by the 
low-S charged-particle process. Another challenge is the recent observation of 
the peculiar Mo isotopic composition of some rare presolar SiC X-grains by \citep{Pellin2000,Pellin2006}, 
which clearly differ from all classical nucleosynthesis processes. \\ 
Let us have a closer look, how the 7 stable Mo (Z=42) isotopes can be synthesized. 
Because of the specific position of Z=42 Mo in the chart of nuclides (see, {\it e.g.} \citet{Pfennig98}), in principle there exists only a narrow nucleosynthesis path for the Mo isotopes in between the stable Zr (Z=40) Zr and Ru (Z=44) isotopes. $^{92,94}$Mo are shielded on the neutron-rich side by their isobars $^{92,94}$Zr. Hence, the p-nuclide $^{92}$Mo can only be produced directly or via ${\beta^+}$-decay from proton-rich isobars like $^{92}$Tc. Apart from its direct synthesis as Mo isotope, the heavier p-nuclide $^{94}$Mo can only be produced by ${\beta^+}$-decay of its proton-rich isobars like $^{94}$Tc and / or by ${\beta^-}$-decay from $^{94}$Nb. Only $^{95}$Mo and $^{97}$Mo can be reached by longer $\beta$-decay chains on both sides of stability. In contrast, the classicial s-only isotope $^{96}$Mo in between is "shielded" on both sides by its stable isobars $^{96}$Zr and $^{96}$Ru. Finally, the two heaviest Mo isotopes, $^{98}$Mo and the classical "r-only" nuclide $^{100}$Mo, are again "shielded" on the 
proton-rich side by their Ru isobars, but can be reached by the full A=98 and 100 ${\beta^-}$-decay chains on the neutron-rich side. \\ 
With this special situation for the different and possibly competing modes of populations, we now can check how the individual Mo isotopes are produced by the $\alpha$-component of the HEW scenario. We find that the lightest stable Mo isotope is only formed directly as $^{92}$Mo in the normal $\alpha$-rich freezeout at low entropies of S$\le$40. For $^{94}$Mo (S$\le$50) and $^{96}$Mo (S$\le$60), besides a predominant direct production, additional minor contributions in the A=94 mass chain come from ${\beta^+}$-decays of $^{94}$Tc and $^{94}$Ru, and in the A=96 chain from $\beta^-$-decay of $^{96}$Nb, respectively. No contributions are predicted by our HEW model from $\beta^-$-decay of $^{94}$Nb and from $\beta^+$-decay of $^{96}$Tc. All other heavier Mo isotopes are no longer produced directly in significant amounts. They are instead predominantly formed as ${\beta^-}$-decay end-products after an increasingly neutron-rich $\alpha$-freezeout at somewhat higher entropies in the range 90$\le$S$\le$150 depending on Y$_e$. The main progenitors of $^{95}$Mo are the ${\beta^-}$-unstable isotopes $^{95}$Y, $^{95}$Zr and $^{95}$Sr, with minor contributions from the even more neutron-rich isobars $^{95}$Rb and $^{95}$Kr. The latter two nuclides are already precursors of $\beta$-delayed neutron ($\beta$dn) emission \citep{pfeiffer02}. Their $\beta$dn-fractions will further $\beta$-decay to stable $^{94}$Zr. For $^{97}$Mo, the main isobaric progenitors are $^{97}$Y, $^{97}$Zr, $^{97}$Sr and the $\beta$dn-precursor $^{97}$Rb. The $\beta$dn-decay of this latter isotope will finally populate $^{96}$Zr. In the A=97 mass chain, in addition small contributions to $^{97}$Mo come from the $\beta$dn-decays of $^{98}$Rb and $^{98}$Kr. The nuclide 
$^{98}$Mo is predominantly formed by the progenitor $^{98}$Sr and to a minor extent by the two $\beta$dn-precursors $^{98}$Rb and $^{98}$Kr. Further small contributions come from the $\beta$dn-decays of $^{99}$Rb and $^{99}$Kr. Finally, the production of the "r-only" isotope $^{100}$Mo within the A=100 mass chain originates exclusively from the neutron-rich progenitor $^{100}$Sr; only small contributions come from very neutron-rich $\beta$dn-precursors of A=101. \\
\begin{figure}[h]
\begin{center}
\centerline{\includegraphics[width=78.mm, angle=0]{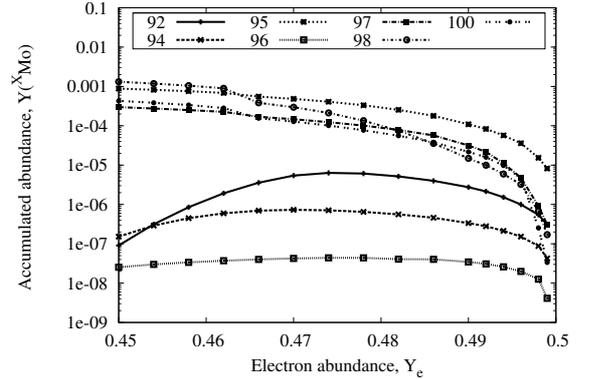}}
\caption{Isotopic Mo abundances Y($^x$Mo) produced by the HEW $\alpha$-process at an expansion velocity of V$_{exp}$=7500 km/s as a function of electron abundance in the range 
0.450$\le$Y$_e$$\le$0.498. The symbols for the Mo isotopes are given in the upper part of the figure. For a discussion of the predicted abundance trends with Y$_e$, see text. }
\label{fig2}
\end{center}
\end{figure}
The isotopic Mo abundances (Y($^x$Mo)) predicted by the charged-particle ($\alpha$-) component of our HEW model as a function of the electron abundance in the range 0.450$\le$Y$_e$$\le$0.498 (at V$_{exp}$=7500 km/s) are shown in Figure~\ref{fig2}. For the whole Y$_e$-range, $^{96}$Mo has the lowest yield of all Mo nuclei, followed by $^{94}$Mo and $^{92}$Mo. The abundances of $^{95,97,98,100}$Mo are considerably higher and lie rather close together. They exhibit a smoothly decreasing pattern with increasing electron abundance up to Y$_e$$\simeq$0.490; for higher Y$_e$ the drop of their abundances becomes more pronounced and reaches -- except for $^{95}$Mo -- the low values of $^{92,94}$Mo. This drop in Y($^x$Mo) in the range 0.490$\le$Y$_e$$\le$0.498 seems to be a general signature for all Mo isotopes. Whereas the abundance patterns of $^{94}$Mo and $^{96}$Mo are rather flat between Y$_e$=0.450 and about 0.490, $^{92}$Mo is the only nuclide which shows a more curved slope with its highest abundance around Y$_e\simeq$0.475. \\ 
In Figur~\ref{fig3}, we show the isotopic abundance {\it ratios} of $^x$Mo/$^{97}$Mo as a function of  Y$_e$. Here we see that the ratios for the heavier s+r isotopes $^{95,97}$Mo are quite flat over the whole Y$_e$-range. The "r-only" nuclide $^{100}$Mo shows a similar behavior up to about Y$_e$$\simeq$0.490; then, the $^{100}$Mo/$^{97}$Mo ratio increases towards Y$_e$=0.498. In contrast, the $^x$Mo/$^{97}$Mo ratios of the two "p-only" isotopes $^{92,94}$Mo and the "s-only" nuclide $^{96}$Mo are smoothly increasing over the whole Y$_e$ range, with a steeper rise in their slopes at about Y$_e$$\ge$0.490. \\ 
\begin{figure}[h]
\begin{center}
\centerline{\includegraphics[width=78.mm, angle=0]{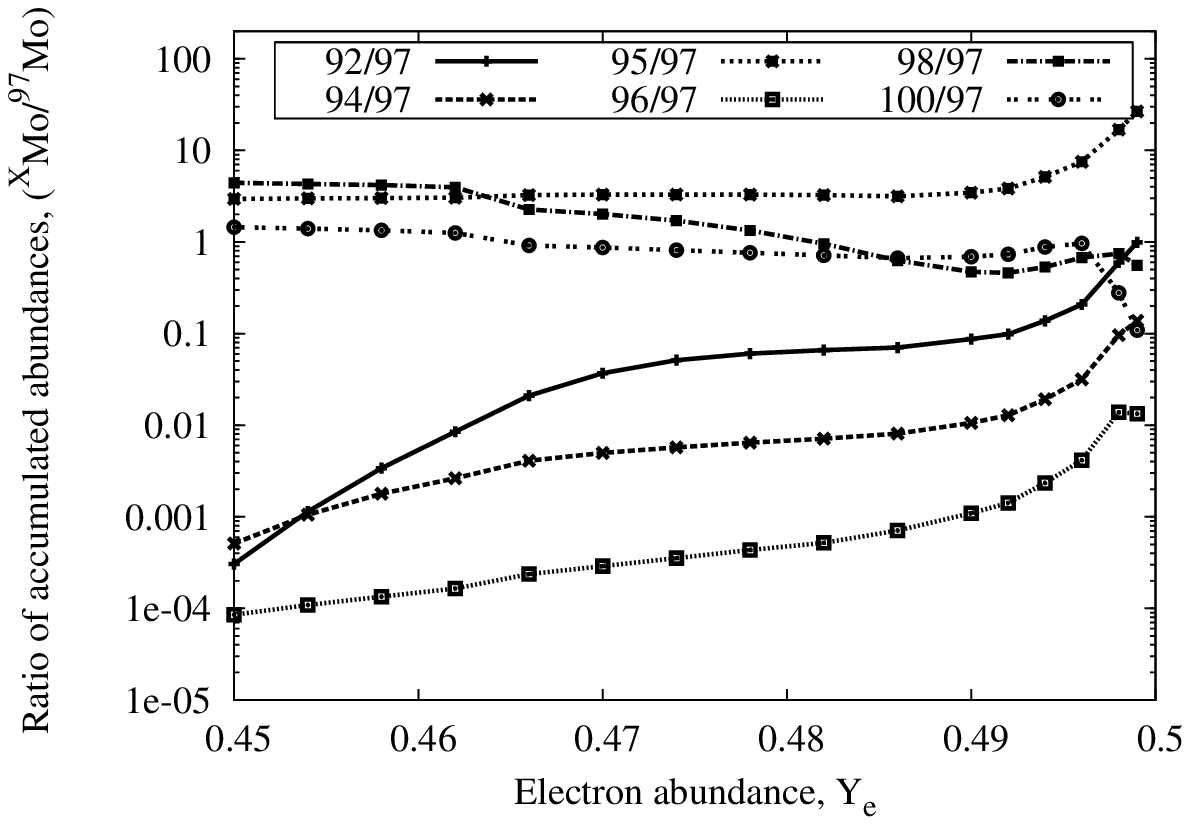}}
\caption{Isotopic abundance ratios of $^x$Mo/$^{97}$Mo obtained for the HEW $\alpha$-process at an expansion velocity of V$_{exp}$=7500 km/s as a function of electron abundance in the range 0.450$\le$Y$_e$$\le$0.498. The symbols for the different isotopic ratios are given in the uppert part of the figure. For a discussion of the predicted trends with Y$_e$, see text.}
\label{fig3}
\end{center}
\end{figure} 

As already mentioned in the introduction, still today the origin of the two p-nuclei 
$^{92}$Mo and $^{94}$Mo is considered to be {\it "one of the great outstanding 
mysteries in nuclear astrophysics"} \citep{Fisker}. Based on the initial ideas 
developed already 30 to 50 yeras ago (see, e.g. 
\citet{b2fh,Arnould76,WoosleyHoward}),
more recently several astrophysical scenarios 
have been investigated in this special context, including for example a p-process based 
on photodisintegration of heavy elements produced by s- and r-processes \citep{ArnouldGoriely}, 
 core-collapse SNe with neutrino-wind, $\gamma$-process and $\nu$p- 
scenarios (see, e.g. \citet{Hoffman96,Rauscher02,Frohlich,Fisker}),
EC SNe \citep{Wanajo}, and X-ray bursts 
(see, e.g. \citep{Schatz89,Weinberg06}). However, none of 
these models has been able to reproduce consistently the high yields and the SS isotopic 
ratio of these two p-isotopes. Finally, also the neutron-capture "burst" 
model of \citep{Meyer}, which has originally been developed to explain 
the Mo isotopic composition in presolar SiC grains, fails to reproduce the SS ratio of 
$^{92}$/$^{94}$Mo.  \\
To summarize this part, we conclude that the low-S charged-particle component of our 
HEW model co-produces all 7 Mo isotopes. With the above S -- Y$_e$ superpositions, 
the SS isotopic ratio of $^{92}$Mo/$^{94}$Mo is reproduced, whereas all other abundance 
ratios are different from the standard s-process and r-process values. While the yields of 
the isotopes $^{92,94,96}$Mo have reached saturation already well below S$_{max}$ 
(with \phantom{-} Y$_n$/Y$_{seed}$=1.0), the production of the heavier, more neutron-rich 
nuclides $^{97,98,100}$Mo continues to higher entropies, reaching the "weak" r-process 
component. \\

Last but not least, we want to present preliminary results of our HEW model 
which offers a new explanation of the puzzling Mo isotopic pattern in presolar 
SiC X-grains recently discovered by \citep{Pellin2000,Pellin2006}. This pattern 
clearly differs from that derived from either a pure s-process or a classical 
r-process. So far, possible nucleosynthesis implications have only been 
successfully analyzed by \citep{Meyer} within their neutron-capture 
scenario in shocked He-rich matter in an exploding massive star, among 
cosmochemists commonly cited as  "neutron-burst" model. \\
This rather complex model starts with a SS "seed" composition (thus initially 
containing already the "correct" SS abundances of the p-nuclei $^{92}$Mo and 
$^{94}$Mo), which is then exposed to a weak neutron fluence in order to mimic weak s-process 
conditions during the pre-SN phase. The weak-s ashes act as secondary seed composition, which 
are then suddenly heated to T$_9$=1.0. During expansion and cooling on a 10 s hydrodynamic 
timescale, a neutron density from ($\alpha$,n) reactions of about 10$^{17}$ n/cm$^{3}$ is 
created for about 1 s.  Neutron density and burst duration are "finetuned" to transform 
the secondary seed composition in the Y to Mo region, so that simultaneously a large mass 
fraction of $^{96}$Zr is obtained, whereas all isotopes between $^{92}$Mo and $^{97}$ Mo are strongly 
and $^{98,100}$Mo are slightly depleted. With this kind of modulation of the seed abundances, 
the "neutron-burst" model 
yields good overall agreement with the Mo isotopic pattern of the SiC X-grains. \\
As already mentioned above, with our HEW model we have found that the isotopic 
abundances of the two "p-only" nuclides $^{92,94}$Mo and the "s-only" isotope 
$^{96}$Mo are already "saturated" well below S$_{max}$ in the charged-particle 
($\alpha$-) component. Under these low entropy conditions (5$\le$S$\le$70), and 
within the correlated electron fractions in the range 0.456$\le$Y$_e$$\le$0.460, we 
indeed obtain a consistent picture reproducing the SiC X-grain pattern. \\
Cosmochemists conventionally compare their measured isotopic abundance ratios 
with model predictions in terms of three-isotope mixing correlations, in order to 
describe the nucleosynthetic origin of their circumstellar grain material. Thereby, 
a best match of the model results to the grain data is deduced from mixing lines 
between the SS composition and the pure nucleosynthesis value (see, as an example 
the three-isotope plot of $^{100}$Mo/$^{97}$Mo versus $^{96}$Mo/$^{97}$Mo in  
Fig. 1 of \citet{Marhas2007}). In Table~\ref{table3} we compare the end members of 
the mixing lines of the SiC grain data with the predictions of the "neutron-burst" model 
and our HEW $\alpha$-nucleosynthesis component with the above S -- Y$_e$ 
superpositions. It is evident from this table, that our HEW approach of synthesizing all 7 Mo 
isotopes in the presolar SiC grains by a primary charged-particle process is (at least) 
an alternative to the secondary production by the "neutron-burst" model. However, 
an advantage of our approach may in principle be the agreement with the astronomical 
observations of the charged-particle nucleosynthesis mode of the light trans-Fe elements 
in UMP halo stars.   \\
\begin{table*}[t]
\begin{center}
\caption{Molybdenum isotopic abundance ratios ($^x$Mo/$^{97}$Mo) as conventionally used in 
cosmochemistry for three-isotope mixing correlations to describe the nucleosynthetic origin 
of presolar silicon carbide (SiC) grains (see, e.g. \citet{Marhas2007}). Comparison 
of recent data deduced from the 8 SiC X-grains measured by \citep{Pellin2000,Pellin2006} with 
the predictions of the "neutron-burst" model of \citep{Meyer} and the 
low-entropy (S$<$70) charged-particle ($\alpha$-)component of our HEW model.
}
\label{table3}
\begin{tabular}{rrrr}
\hline
&\multicolumn{3}{c}{Isotopic abundance ratios} \\
 $^x$Mo/$^{97}$Mo    &    SiC X-grains &     This work  &    "n-burst" model \\
\hline
$^{92}$Mo/$^{97}$Mo   &     $<$ E-2      &         4.1 E-3  &           1.43 E-3 \\
$^{94}$Mo/$^{97}$Mo   &     $<$ E-2      &         6.3 E-3  &           3.27 E-4 \\ 
$^{95}$Mo/$^{97}$Mo   &       2.1     &      3.12       &          1.539 \\
$^{96}$Mo/$^{97}$Mo   &      0.12    &   4.77 E-2    &        1.02 E-2  \\
$^{98}$Mo/$^{97}$Mo   &      1.2      &     0.950      &           0.382  \\
$^{100}$Mo/$^{97}$Mo &      0.25    &     0.225      &        9.55 E-2  \\
\hline
\end{tabular}
\end{center}
\end{table*}
 
\section{Summary and conclusion} 
We have shown in a large-scale parameter study that the high-entropy wind (HEW) scenario of 
type II supernovae can co-produce the light p-, s- and r-process isotopes between Zn (Z=30) 
and Ru (Z=44) at electron abundances in the range 0.450$\le$Y$_e$$\le$0.498 and low entropies 
of S$\le$100--150. Under these conditions, the light trans-Fe elements are produced 
in a charged-particle ($\alpha$-)process, including all p-nuclei 
up to $^{96,98}$Ru. In our model, no initial SS, s- or r-process seed composition  is invoked; 
hence, this nucleosynthesis component is primary. These results provide a means to substantially 
revise the abundance estimates of different primary and secondary nucleosynthesis processes in 
the historical "weak-s" and "weak-r" process regions. Chosing the Mo isotopic chain as a particularly 
interesting example, we have found that our HEW model can account for the simultaneous production of all 7 
stable Mo nuclides, from p-only $^{92}$Mo, via s-only $^{96}$Mo up to r-only $^{100}$Mo. 
Furthermore, we have shown that our model is able to reproduce the SS abundance ratio of the two 
p-nuclei $^{92}$Mo and $^{94}$Mo. 
Finally, the likely nucleosynthesis origin of the peculiar Mo isotopic composition
of the presolar SiC X-grains measured by \citep{Pellin2000,Pellin2006} has been determined. \\
To obtain more quantitative answers to questions concerning the astrophysical site(s) of the light trans-Fe 
elements will require on the one hand more and higher-quality observational data and on the other 
hand more realistic hydrodynamical nucleosynthesis calculations. In particular, it has to be studied 
in detail how severe overproductions of the SS abundances between Sr (Z=38) and Cd (Z=48) can 
be avoided when combining the partly high yields of all presently favored contributing processes 
for the trans-Fe elements, i.e. the early $\nu$p-process \citep{Frohlich}, the subsequent HEW 
charged-particle process after normal and neutron-rich $\alpha$-freezeout, possible ejecta from X-ray bursts
\citep{Weinberg06} and the new strong s-process 
predicted to occur in massive stars at halo metallicity \citep{Pignatari}.

\section*{Acknowledgments}
We thank R. Gallino, Y. Kashiv, U. Ott and F.-K. Thielemann for helpful discussions. 
K.F. acknowledges financial support from the Joint Institute for Nuclear Astrophysics 
(JINA; PHY 02-16783) and from the Max-Planck-Institut f\"ur Chemie during his stay 
in Mainz.

\end{document}